\documentclass[aip,apl,preprint,superscriptaddress,showpacs]{revtex4-1}%%% <===== single-column style

\usepackage{graphicx}
\usepackage{amsmath}

\begin{document}

\title{A generalized magnetic refrigeration scheme}

\author{Ryo Tamura}
\email[]{TAMURA.Ryo@nims.go.jp}
\affiliation{International Center for Young Scientists, National Institute for Materials Science, 1-2-1, Sengen, Tsukuba, Ibaraki 305-0047, Japan}

\author{Takahisa Ohno}
\email[]{OHNO.Takahisa@nims.go.jp}
\affiliation{Computational Materials Science Unit, National Institute for Materials Science, 1-2-1 Sengen, Tsukuba, Ibaraki 305-0047, Japan}

\author{Hideaki Kitazawa}
\email[]{KITAZAWA.Hideaki@nims.go.jp}
\affiliation{Quantum Beam Unit, National Institute for Materials Science, 1-2-1 Sengen, Tsukuba, Ibaraki 305-0047, Japan}

\date{\today}

\begin{abstract}

We have investigated the magnetocaloric effects in antiferromagnets and compared them with those in ferromagnets using Monte Carlo simulations.
In antiferromagnets, the magnetic entropy reaches a maximum value at a finite magnetic field when the temperature is fixed below the N\'eel temperature.
Using the fact, we proposed a protocol for applying magnetic fields to achieve the maximum efficiency for magnetic refrigeration in antiferromagnets.
In particular, we found that at low temperatures, antiferromagnets are more useful for magnetic refrigeration than ferromagnets. 

\end{abstract}

%\pacs{test}

%\keywords{}

\maketitle

%%%%%%%%%%%%%%%%%%%%%%%%%%%%%%%%%%%%%
%INTRODUCTION
%%%%%%%%%%%%%%%%%%%%%%%%%%%%%%%%%%%%%

%\section{Introduction}

The importance of magnetic refrigeration has been well recognized.\cite{Pecharsky-1999,Novotny-1999,Pecharsky-2001,Tishin-2003,Zhitomirsky-2003,Gschneidner-2005,Sakai-2009,Zvere-2010,Oliveira-2010,Mamiya-2010,Franco-2012,Yonezawa-2013}
In conventional ferromagnets and paramagnets, when the applying magnetic field is turned off,
the magnetic entropy increases.
Thus, these magnetic materials absorb an amount of heat associated with the magnetic entropy change, and the temperature drops.
This is the magnetocaloric effect (MCE), which can be applied in magnetic refrigeration.
From a viewpoint of magnetic refrigeration,
magnetic materials which exhibit a large MCE under a small applied magnetic field are regarded as good materials.
The large MCE was observed around the Curie temperature in ferromagnets using the conventional protocol for applying magnetic fields where the magnetic field is varied from a finite value to zero.
Thus, ferromagnets are likely to be suitable materials for magnetic refrigeration.\cite{Fisher-1973,Brown-1976,Hashimoto-1981,Pecharsky-1997,Hu-2001,Wada-2001,Wada-2002,Tegus-2002,Fujieda-2002,Fujita-2003,Chikazumi-2009,Mizumaki-2013}

In recent years, it was reported in many experimental researches that antiferromagnets are also useful for magnetic refrigeration.
\cite{Samanta-2007a,Samanta-2007b,Hu-2008,Li-2009a,Li-2009b,Chen-2009,Naik-2011,Kim-2011}
There are two features on the magnetic entropy reported in these experimental researches when the conventional protocol for applying magnetic fields is used.
First one is that the inverse MCE, where the magnetic entropy decreases when the applying magnetic field is turned off, was observed below the N\'eel temperature.
Second one is that a large MCE similar to that in ferromagnets was obtained around the N\'eel temperature.
These complicated behaviors of the magnetic entropy were not observed in conventional ferromagnets and paramagnets.
Thus, in order to understand the potential of antiferromagnets for magnetic refrigeration,
the microscopic features of the MCE in antiferromagnets, which are not well understood, should be investigated.

%%%%%%%%%%%%%%%%%%%%%%%%%%%%%%%%%%%%%
%PURPOSE
%%%%%%%%%%%%%%%%%%%%%%%%%%%%%%%%%%%%%

%\section{Purpose}

The purpose of this letter is to present the microscopic features of the MCE in antiferromagnets by using Monte Carlo simulations.
We show that the magnetic entropy reaches a maximum value at a finite magnetic field when the temperature is fixed below the N\'eel temperature.
Based on the fact, we propose a protocol for applying magnetic fields to achieve the maximum efficiency for magnetic refrigeration in antiferromagnets.
By using our proposed protocol, we find that antiferromagnets exhibit a large magnetic entropy change in the ordered phase below the N\'eel temperature rather than around the N\'eel temperature.
Furthermore,
we show that at low temperatures, antiferromagnets are more useful for magnetic refrigeration than ferromagnets.

%%%%%%%%%%%%%%%%%%%%%%%%%%%%%%%%%%%%%
%MODEL
%%%%%%%%%%%%%%%%%%%%%%%%%%%%%%%%%%%%%

%\section{Model and calculation method}

\begin{figure*}[t]
\begin{center}
\includegraphics[scale=1.0]{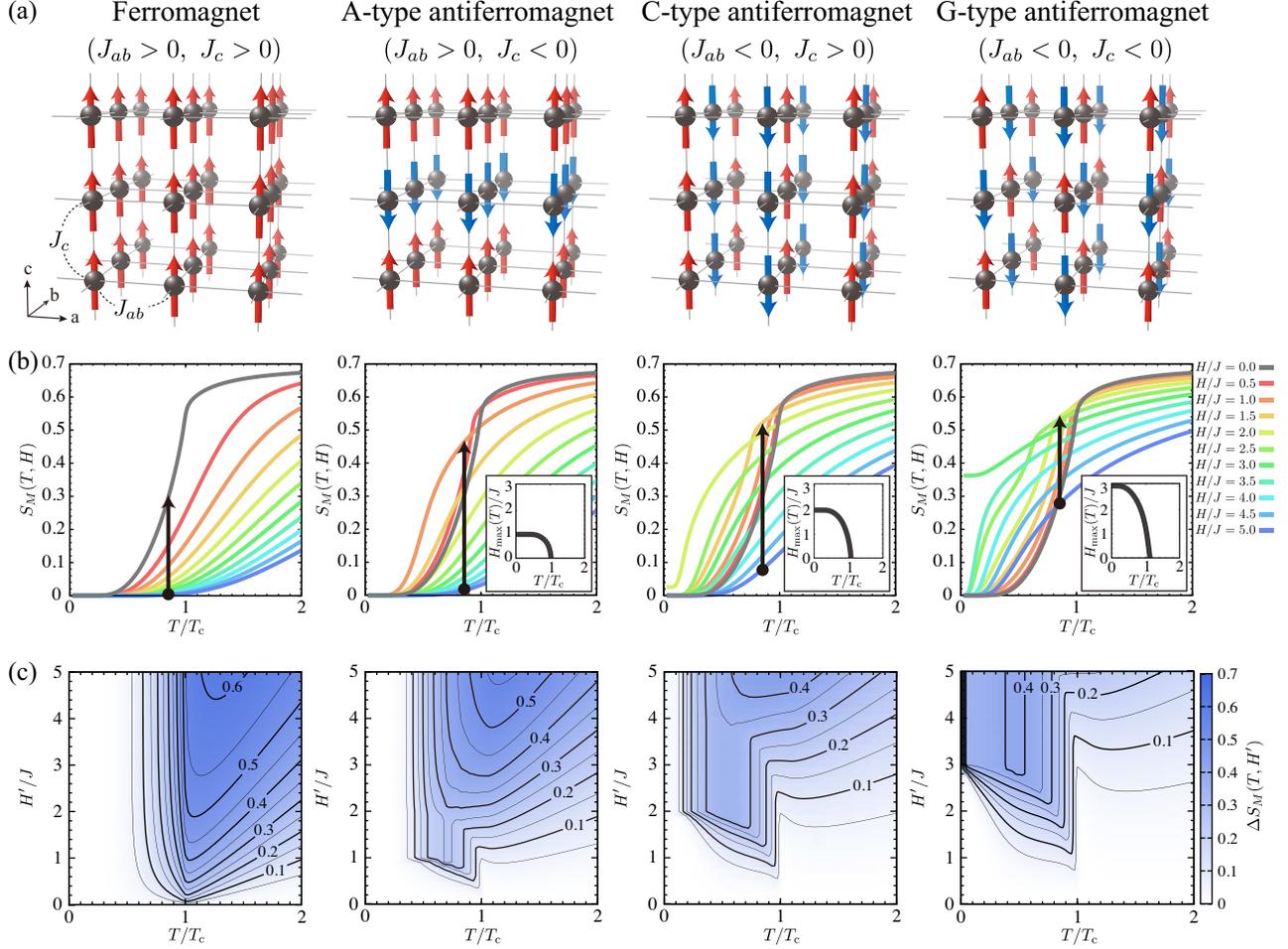} 
\end{center}
\caption{\label{fig:entropy}
(Color online)
(a) Schematic of ordered magnetic structures in the Ising model on a simple cubic lattice.
The signs of the magnetic interactions in the $ab$-plane, $J_{ab}$, and along the $c$-axis, $J_c$, where each magnetic structure is in the ground state are shown.
(b) Temperature $T/T_\text{c}$ dependence of the magnetic entropy per spin $S_M (T,H)$ for $L=16$ under several magnetic fields.
Arrow is the isothermal demagnetization process in which the magnetic entropy change becomes a maximum from $T/T_\text{c}=0.85$ and $H/J=5$.
Inset shows $T/T_\text{c}$ dependence of $H_{\rm max} (T)/J$ at which $S_M (T,H)$ reaches a maximum value.
(c) Contour map of the magnetic entropy change $\Delta S_M(T,H')$ defined in Eq.~(\ref{eq:delta_S}).
}
\end{figure*}

To explore the relation between the ordered magnetic structure and MCE in a unified way, 
we study MCEs in the Ising models on a simple cubic lattice with the periodic boundary condition.
Here, let $N=L \times L \times L$ be the number of spins in a simple cubic lattice, where $L$ is the linear dimension.
The model Hamiltonian is defined by 
\begin{align}
\mathcal{H} = - J_{ab} \sum_{\langle i,j \rangle_{ab}} s_i^z s_j^z 
                      - J_{c} \sum_{\langle i,j \rangle_{c}} s_i^z s_j^z
                      - H \sum_{i} s_i^z,
                      \ \ \ \ \ s_i^z = \pm \frac{1}{2}, \label{eq:model}
\end{align}
where the first and second sums are over nearest-neighbor sites in the $ab$-plane and along the $c$-axis, respectively,
and $J_{ab}$ and $J_c$ represent magnetic interactions.
Furthermore, $H$ denotes a uniform magnetic field along the $z$-axis, where $g$-factor and the Bohr magneton $\mu_\text{B}$ are set to unity.
When the sign of the magnetic interaction is positive, the magnetic interaction is ferromagnetic,
whereas the magnetic interaction is antiferromagnetic when the sign is negative.
For simplicity, we consider the case that the absolute values of $J_{ab}$ and $J_{c}$ are the same,
that is $J:=|J_{ab}|=|J_{c}|$, where $J$ is the energy unit.
At zero magnetic field ($H/J=0$),
the system exhibits a second-order phase transition at the critical temperature $T_\text{c}/J=1.127\cdots$\cite{Ferrenberg-1991} independent of the signs of $J_{ab}$ and $J_c$,
where the Boltzmann constant $k_\text{B}$ is set to unity.
In this letter, 
we focus on four combinations of interactions (ordered magnetic structures): 
(i) $J_{ab}>0$, $J_{c}>0$ (ferromagnet),
(ii) $J_{ab}>0$, $J_{c}<0$ (A-type antiferromagnet),
(iii) $J_{ab}<0$, $J_{c}>0$ (C-type antiferromagnet),
and (iv) $J_{ab}<0$, $J_{c}<0$ (G-type antiferromagnet).
Figure~\ref{fig:entropy} (a) shows the ordered magnetic structure in the ground state for each case.
Note that although the net magnetization is zero in each antiferromagnetic structure,
the number of antiferromagnetic interactions at each site is different.
Namely, the numbers of antiferromagnetic interactions at each site in the A-type, C-type, and G-type antiferromagnets are $2$, $4$, and $6$, respectively.

We use the Wang-Landau method\cite{Wang-2001a, Wang-2001b,Lee-2006} in Monte Carlo simulations to calculate the temperature $T$ and $H$ dependence of the magnetic entropy with high accuracy.
In the Wang-Landau method,
we use a random walk in the energy space to obtain the absolute density of states.
Then, we can directly calculate the magnetic entropy without integrating magnetization or specific heat.
For details of the Wang-Landau method,
see Refs.~[35,36].
%\cite{Wang-2001b,Lee-2006}

%%%%%%%%%%%%%%%%%%%%%%%%%%%%%%%%%%%%%
%RESULT
%%%%%%%%%%%%%%%%%%%%%%%%%%%%%%%%%%%%%

%\section{Result}

Figure~\ref{fig:entropy} (b) shows the magnetic entropy per spin $S_M(T,H)$ as a function of $T/T_\text{c}$ for $H/J=0$ - $5$ when the lattice size is $L=16$.
Here, the unit of $S_M(T,H)$ is the Boltzmann constant $k_\text{B}$,
and thus the magnetic entropy per mol is obtained by $k_\text{B} N_\text{A} S_M (T,H) \ [\text{J}/\text{mol} \ \text{K}]$ where $N_\text{A}$ is the Avogadro's number.
Furthermore, since the spin degree of freedom is two in the Ising model, 
the maximum value of $S_M(T,H)$ is $\ln 2 = 0.693\cdots$.
When $H/J=0$, the results do not depend on the magnetic structure.
The magnetic entropies for $L=8$, $12$, and $16$ overlap within the line width in Fig.~\ref{fig:entropy} (b),
and thus the lattice size dependence of the magnetic entropy is negligibly small.
Therefore, we use a lattice size of $L=16$ throughout this paper.
In the ferromagnet,
magnetic entropy increases as the magnetic field decreases at any temperature.
The same behaviors are obtained in the paramagnetic phase of antiferromagnets above $T_{\rm c}/J$.
In contrast,
below $T_{\rm c}/J$ for antiferromagnets,
there is the case that the magnetic entropy decreases as the magnetic field decreases, which is the origin of the inverse MCE.
Because the magnetic field does not favor the spin configuration in the antiferromagnetically ordered phase,
the antiferromagnetic state is destroyed by applying the magnetic field.
This behavior suggests that in antiferromagnets,
there is a finite magnetic field $H_\text{max} (T)$ at which the magnetic entropy reaches a maximum value when the temperature is fixed below $T_\text{c}/J$.
The inset of Fig.~\ref{fig:entropy} (b) shows $T/T_\text{c}$ dependence of $H_\text{max} (T)$ for each antiferromagnetic structure.
Note that $H_\text{max} (T)$ is always zero in the ferromagnet as mentioned above.
The behavior of $H_\text{max} (T)$ below $T_\text{c}/J$ in antiferromagnets indicates that we should use a new protocol for applying magnetic fields to obtain the maximum magnetic entropy change.
That is, the magnetic field should be varied from a finite value to $H_\text{max} (T)$ instead of zero below $T_\text{c}/J$ in antiferromagnets.
For example, suppose we consider the isothermal demagnetization process from $T/T_\text{c}=0.85$ and $H/J=5$.
The processes in which the magnetic entropy change becomes a maximum are denoted by arrows in Fig.~\ref{fig:entropy} (b).
In the ferromagnet, when the magnetic field is turned off, the maximum magnetic entropy change is obtained.
In contrast, in each antiferromagnet,
the magnetic entropy change obtained from our proposed protocol where $H/J$ is varied from $5$ to $H_\text{max} (T)/J$ is larger than that obtained from the conventional protocol where $H/J$ is varied from $5$ to $0$.

Next, we consider a temperature region in which a large magnetic entropy change is obtained in each magnetic structure.
Here, we define the magnetic entropy change by
\begin{align}
\Delta S_{M} (T,H') := \text{max} &\{ S_{M} (T,H)|H \le H' \} - \text{min} \{ S_{M} (T,H)|H \le H' \}, \label{eq:delta_S}
\end{align}
where $H'$ is the maximum value of applied magnetic field.
$\Delta S_{M} (T,H')$ indicates the maximum magnetic entropy change regardless of a protocol for applying magnetic fields when the magnetic field $H$ $(\le H')$ is applied.
Note that when $H' \ge H_\text{max}(T)$, $\text{max} \{ S_{M} (T,H)|H \le H' \} = S_M(T,H_\text{max} (T))$.
Figure~\ref{fig:entropy} (c) shows $T/T_\text{c}$ and $H'/J$ dependence of $\Delta S_{M} (T,H')$.
In the ferromagnet,
$\Delta S_{M} (T,H')$ has a large value around $T_\text{c}/J$.
Thus, the ferromagnet exhibits a large magnetic entropy change around the Curie temperature.
However, in the ordered phase below $T_\text{c}/J$,
the magnetic entropy change is exceedingly small.
In contrast, in antiferromagnets,
$\Delta S_{M} (T,H')$ becomes large below $T_\text{c}/J$.
This indicates that antiferromagnets can exhibit a large magnetic entropy change in the ordered phase below the N\'eel temperature rather than around the N\'eel temperature.
Moreover, the temperature region in which $\Delta S_{M} (T,H')$ has a large value moves towards lower temperature as increasing the number of antiferromagnetic interactions.

\begin{figure}[t]
\begin{center}
\includegraphics[scale=1.0]{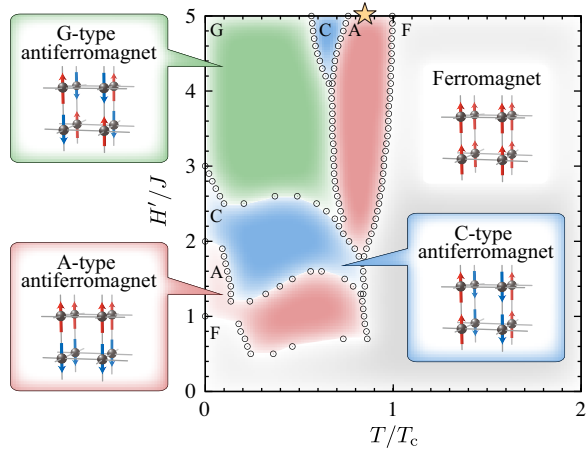} 
\end{center}
\caption{\label{fig:type}
(Color online)
A set of regions such that each magnetic structure exhibits a larger magnetic entropy change than other three magnetic structures.
The star indicates the parameter used in Fig.~\ref{fig:hmax}.
}
\end{figure}

Based on the results of $\Delta S_{M} (T,H')$,
Fig.~\ref{fig:type} shows a region in which each magnetic structure is most suited for magnetic refrigeration from among the four types of magnetic structures.
That is, in each region, a drawn magnetic structure exhibits a larger magnetic entropy change than other three magnetic structures. 
The obtained value of magnetic entropy change in each region can be known from corresponding contour map of $\Delta S_M (T,H')$ shown in Fig.~\ref{fig:entropy} (c). 
Figure~\ref{fig:type} indicates that the ferromagnet is always suited for magnetic refrigeration at high temperatures above $T_\text{c}/J$. 
In contrast, at low temperatures below $T_\text{c}/J$, there is a wide region where antiferromagnets are more useful for magnetic refrigeration than the ferromagnet.

\begin{figure}[t]
\begin{center}
\includegraphics[scale=1.0]{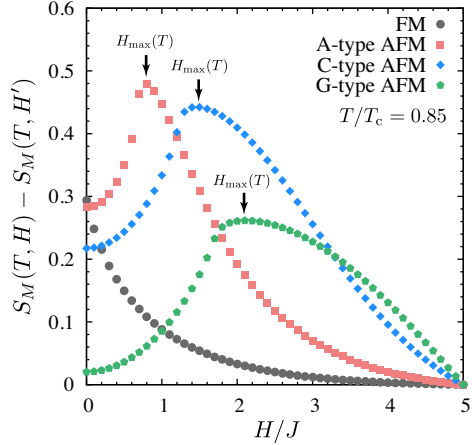} 
\end{center}
\caption{\label{fig:hmax}
(Color online)
Magnetic field $H/J$ dependence of the difference between magnetic entropies $S_M(T,H)-S_M(T,H')$ at $T/T_\text{c}=0.85$ for ferromagnet (FM) and three types of antiferromagnets (AFMs).
The maximum applied magnetic field is $H'/J=5$.
}
\end{figure}

We showed that at low temperatures, antiferromagnets exhibit a larger magnetic entropy change than the ferromagnet when the magnetic field is varied from a finite value to $H_\text{max} (T)$ instead of zero.
Below $T_\text{c}/J$ for antiferromagnets, $H_\text{max} (T)$ is nonzero value shown in the inset of Fig.~\ref{fig:entropy} (b),
and thus $H_\text{max} (T)$ must be known to execute our proposed protocol for applying magnetic fields.
Here, we present a method by which $H_\text{max} (T)$ can be easily obtained for antiferromagnets below $T_\text{c}/J$.
Note that $H_\text{max} (T)$ is always zero above $T_\text{c}/J$ as mentioned above.
Suppose we calculate the difference between magnetic entropies $S_M(T,H)-S_M(T,H')$ as a function of $H$ $(\le H')$ at a fixed temperature.
In antiferromagnets, there should be a peak in the difference when $H' \ge H_\text{max}(T)$,
and the peak position is $H_\text{max} (T)$.
For example,
$H/J$ dependence of $S_M(T,H)-S_M(T,H')$ with $H'/J=5$ at $T/T_\text{c}=0.85$ of the Ising models defined by Eq.~(\ref{eq:model}) is shown in Fig.~\ref{fig:hmax}.
The position of the parameters is indicated in Fig.~\ref{fig:type}.
The value of $S_M(T,H)-S_M(T,H')$ at $H=H_\text{max} (T)$ represents $\Delta S_{M} (T,H')$ defined by Eq.~(\ref{eq:delta_S}) when our proposed protocol is used.
In this case, 
it is clear that the A-type antiferromagnet is the most suitable as a magnetic refrigeration material.
In contrast,
the value of $S_M(T,H)-S_M(T,H')$ at $H=0$ is the magnetic entropy change using the conventional protocol where the magnetic filed is varied from $H'$ to zero.
If the conventional protocol is used, the ferromagnet is regarded as the most suitable as a magnetic refrigeration material.
The magnetic entropy change obtained from our proposed protocol increases by 170 \% (resp. 200 \% and 1280 \%) compared with that obtained from the conventional protocol in the A-type antiferromagnet (resp. C-type and G-type antiferromagnets).
The method to obtain $H_\text{max} (T)$ can be used with the thermodynamic formula:
\begin{align}
S_M(T,H)-S_M(T,H') = \int_{H'}^H \left( \frac{\partial M}{\partial T} \right)_{H''} d H'',
\end{align}
where $M$ is the magnetization, and the integrating interval is $[H',H]$.
This means that only the magnetization process under various temperatures is required.
Thus, this method can be performed by data which were already obtained in experimental researches on magnetic refrigeration.
Moreover, the value of $S_M(T,H)-S_M(T,H')$ can be also estimated by the specific heat.

%%%%%%%%%%%%%%%%%%%%%%%%%%%%%%%%%%%%%
%CONCLUSION
%%%%%%%%%%%%%%%%%%%%%%%%%%%%%%%%%%%%%

%\section{Conclusion}
In conclusion,
we demonstrated the microscopic features of the magnetocaloric effects in the ferromagnetic and antiferromagnetic Ising models by Monte Carlo simulations based on the Wang-Landau method.
In antiferromagnets, the magnetic entropy reaches a maximum value at a finite magnetic field $H_\text{max} (T)$ when the temperature is fixed below the N\'eel temperature.
Thereby, in order to obtain the maximum magnetic entropy change below the N\'eel temperature, 
the magnetic field should be varied from a finite value to $H_\text{max} (T)$ instead of zero.
By using this protocol,
we found that antiferromagnets exhibit a large magnetic entropy change in the ordered phase below the N\'eel temperature rather than around the N\'eel temperature.
We also showed that antiferromagnets are more useful for magnetic refrigeration than ferromagnets at low temperatures.
In non-ferromagnetic materials,
the ordered state is destroyed by applying the magnetic field,
and there should be a finite magnetic field $H_\text{max} (T)$ at which the magnetic entropy reaches a maximum value.
Thus,
our proposed protocol for applying magnetic fields can be widely applied to non-ferromagnetic materials to achieve a maximum efficiency for magnetic refrigeration.

%%%%%%%%%%%%%%%%%%%%%%%%%%%%%%%%%%%%%
%ACKNOWLEDGMENTS
%%%%%%%%%%%%%%%%%%%%%%%%%%%%%%%%%%%%%

%\section*{Acknowledgment}

We thank Kenjiro Miyano and Shu Tanaka for useful comments and discussions.
R.T. and H. K. were partially supported by a Grand-in-Aid for Scientific Research (C) (Grant No. 25420698).
In addition, R. T. was partially supported by National Institute for Materials Science.
The computations in the present work were performed on super computers at the Supercomputer Center, Institute for Solid State Physics, University of Tokyo and National Institute for Materials Science.

\end{document}